
%
\overfullrule 0mm
\input harvmac.tex
\input amssym.def
\input amssym.tex
\input epsf.tex

\def\slh{\widehat{sl}}

\def\CC{{\cal C}}

\def\CH{{\cal H}}
\def\CI{{\cal I}}

\def\CT{{\cal T}}
\def\CV{{\cal V}}\def\CV{{\cal V}}

\def\bi{\bar i}

\def\bj{\bar j}
\def\bn{{\bf n}}
\def\bbn{\bar{\bf n}}

\def\btq{{\tilde{q}}}  
\def\bw{\bar w}
\def\bz{\bar z}

\def\bL{\bar L}
\def\bT{\bar T}

\def\tq{\tilde q}
\def\ttau{\tilde{\tau}}

\def\tN{{\widetilde N}}
\def\tV{{\widetilde V}}
\def\tG{{\widetilde G}}
\def\tCV{{\widetilde {\cal V}}}
\def\za{\alpha} \def\zb{\beta}
\def\zg{\gamma} \def\zd{\delta}

\def\tr{{\rm tr\,}}

\def\omit#1{{}}
\def\plb#1#2#3{Phys. Lett. {\bf B #1} (#2) #3}
\def\npb#1#2#3#4{Nucl. Phys. {\bf B #1} [FS#2] (#3) #4}
\def\cmp#1#2#3{Comm. Math. Phys. {\bf  #1} (#2) #3}
\def\lmp#1#2#3{Lett. Math. Phys. {\bf  #1} (#2) #3}
\def\hepth#1{{\tt hep-th/#1}}
\def\qalg#1{{\tt q-alg/#1}}
\def\mathOA#1{{\tt math.OA/#1}}

\lref\Ocn{A. Ocneanu, 
{\it Paths on Coxeter Diagrams: From Platonic Solids
and Singularities to Minimal Models and Subfactors }
in {\sl Lectures on Operator Theory, }
 {Fields Institute, Waterloo, Ontario, April 26--30, 1995, }{(Notes taken by
S. Goto)}
{Fields Institute Monographies, AMS 1999,}{ Rajarama Bhat et al, eds.};
{\sl Quantum symmetries for $SU(3)$ CFT Models, }
{Lectures at Bariloche Summer School, }
{Argentina, Jan 2000}, to appear 
in AMS Contemporary Mathematics, R. Coquereaux, A. Garcia and
R. Trinchero, eds.}
\lref\BEK{ J. B\"ockenhauer and  D. E. Evans,
\cmp{205}{1999}{183-228} \hepth{9812110};
J. B\"ockenhauer , D. E. Evans and Y. Kawahigashi,
\cmp{208}{1999}{429-487}
\mathOA{9904109}; \cmp{210}{2000}{733-784}
\mathOA{9907149}. } 
\lref\BPPZ{R.E. Behrend, P.A. Pearce, V.B. Petkova and J.-B.
Zuber, \plb{444}{1998}{163-166}, 
\hepth{9809097};
\npb{579}{}{2000}{707-773}, \hepth{9908036}.}
\lref\BSz{G. B\"ohm and K. Szlachanyi, \lmp{200}{1996}{437-456}
\qalg{9509008}}
\lref\Cab{J.Cardy, \npb{275}{17}{1986}{200-218}.}
\lref\Ca{J.Cardy, \npb{270}{16}{1986}{186-204}.}
\lref\OP{W. Orrick and P. Pearce, to be published.}
\lref\Zub{J.-B. Zuber, Phys. Lett. {\bf B 176} (1986) 127-128.}
\lref\vGR{G. von Gehlen and V. Rittenberg, J. Phys. {\bf A19}
 (1986) L625-L629.}
\lref\Ruel{S. Li\'enart, P. Ruelle and O. Verhoeven,
{\sl On discrete symmetries in $su(2)$ and $su(3)$ affine theories and 
related graphs}, \hepth{0007095}. }
\lref\PZtmr{ V.B. Petkova and J.-B. Zuber,
{}PRHEP-tmr2000/038  (Proceedings of the TMR network conference
{\it Nonperturbative Quantum Effects 2000}),
 \hepth{0009219}.}
\lref\PZnew{ V.B. Petkova and J.-B. Zuber,
{\sl The many faces of Ocneanu cells}, in preparation.}
\lref\Reck{A. Recknagel, in preparation. }

\font\Huge=cmbx10 scaled \magstep2

\vbox{\vglue-8mm\baselineskip12pt\hbox{UNN-SCM-M-00-07}\hbox{CERN-TH/2000-322} }
\bigskip\centerline{{\Huge Generalised twisted 
partition functions }}
\bigskip
\centerline{V.B. Petkova
\footnote{${}^*$}{Permanent address:
Institute for Nuclear Research and Nuclear Energy,
72 Tzarigradsko Chaussee,
1784 Sofia, Bulgaria. E-mail: {valentina.petkova@unn.ac.uk}}}
\centerline{School of Computing and Mathematics}
\centerline{University of Northumbria}
\centerline{NE1 8ST Newcastle upon Tyne, UK}
\centerline{and}
\centerline{J.-B. Zuber
\footnote{${}^\flat$}{On leave from: SPhT, CEA Saclay, 91191 Gif-sur-Yvette, France. E-mail:  {zuber@spht.saclay.cea.fr}}}
\centerline{TH Division, CERN, CH-1211 Gen\`eve 23}

\bigskip
\noindent
{\baselineskip9pt {\ninepoint 
We consider the set of partition functions that result from the 
insertion of twist operators compatible with conformal invariance 
in a given 2D  Conformal Field Theory (CFT).  A consistency equation, 
which gives a classification of twists,  is written and solved in 
particular cases. This generalises old results on twisted torus 
boundary conditions, gives a physical interpretation of Ocneanu's 
algebraic construction, and might offer a new route to the study 
of properties of CFT. }}


\newsec{Introduction}
\noindent
The study of possible boundary conditions and of the associated 
finite size effects is known to be a powerful means of investigation of 
critical systems. This is particularly true in two dimensions, where
conformal invariance gives a very restrictive framework. In this note
we discuss a class of twists which may be inserted 
in  a 2D conformal field theory along a non contractible cycle 
(on a cylinder, say), and which are requested to be compatible 
with conformal invariance, in a sense to be defined. It is
 shown that the complete set of partition functions 
depending of these twists and the associated algebras contain 
all the information on the system, its physical spectrum and Operator
Product Algebra. Our discussion is parallel to the one done recently
on boundary conditions on a half plane (or on a strip) \BPPZ. 
Somehow, the latter explored a chiral subsector  of the theory, 
while the present approach reveals all its structure. 
At the same time it  gives a physical realisation of an algebraic
construction proposed  by Ocneanu \Ocn\ and pursued also by B\"ockenhauer,
Evans and Kawahigashi~\BEK. 
\newsec{Twisted boundary conditions}
\noindent
Throughout this paper we make use of the following notations:
We consider a rational conformal field theory (RCFT)
with a chiral algebra ${\goth A}$, (the Virasoro algebra or one of
its extensions), and denote 
$\{\CV_i\}_{i\in \CI}$ the finite set of representations of this
chiral algebra, $\chi_i(q)$, $S_{ij}$, $N_{ij}{}^k$
with $i,j,k\in \CI$ their 
characters, modular matrix and fusion rule multiplicities
given by the Verlinde formula. 

Suppose this RCFT has a 
spectrum in the plane described by  a matrix $Z_{j\bj}$, $j,\bj\in \CI$, 
i.e. a Hilbert space of the form
\eqn\hilbert{
\CH_P=\oplus_{j,\bj\in \CI}\, Z_{j\bj}\, 
\CV_j\otimes \overline{\CV}_{\bj}\,.} 
In order to study this CFT, it is a common practice \Ca\ to 
consider it on a cylinder  of perimeter $T$ with a complex coordinate 
$w$ defined modulo $T$ and to  define a Hamiltonian $H$ by the translation
operator along the $w$ imaginary axis. Then, 
taking a  finite portion of this cylinder
bounded by the circles $\Im m\, w =\pm L$, one may
identify the two boundaries by imposing periodic boundary
conditions along this imaginary direction: 
the CFT is regarded as  living  on a torus, and the partition function
is the trace of the ``time'' evolution operator
$\CT$ associated with the Hamiltonian, $\CT=e^{-2LH}$. 
 Here, however, we shall allow  
the possibility of inserting one (or several) operator(s) $X$ 
inside the trace of this evolution operator. This may be interpreted
as introducing one 
 or several defect lines $\CC$ (`` seams'') in the system,
along  non 
contractible cycles of the cylinder, before closing it into a torus, 
thus resulting into a certain class of ``twisted'' boundary conditions.
The new partition functions in the presence of $X$'s are thus $Z_X=\tr_{\CH_P} 
X \CT$,  $Z_{X,X'}=\tr_{\CH_P} X X' \CT$, etc.
 
The $X$ are not arbitrary:  we insist that these operators commute with 
the energy-momentum tensor $T(w), \bT(\bw)$, or equivalently
with the Virasoro generators
\eqn\commut
{ [L_n,X]=[\bL_n,X]=0 \ .   }  
Since the Virasoro operators are the generators of infinitesimal
diffeomorphisms, this condition says 
that each operator $X$ is invariant under a distorsion of the 
line to which it is attached.  The operator $X$ 
is thus attached to  the homotopy class of the contour $\CC$. 
If the chiral algebra is larger than Vir, there is a similar set of
commutation  relations with the generators of ${\goth A}$, 
whose physical interpretation is however less obvious.


\newsec{Characterising the twists}
\noindent 
What is the most general form of operators from $\CH_P$ to $\CH_P$
commuting with all $L_n$ and $\bL_n$?
 Following a route that proved useful in a different context
\BPPZ, we may first restrict ourselves  to operators intertwining
a pair of components of \hilbert, i.e. mapping some 
$\CV_j\otimes \overline{\CV}_{\bj}$ into $\CV_{j'}\otimes
 \overline{\CV}_{\bj'}$:
irreducibility of the representations $\CV_j$ tells us such an $X$ is 
non trivial only for $j=j'$, $\bj=\bj'$. 
If the multiplicity $Z_{j\bj}$ is 1, it follows that $X$ must be 
proportional to the projector $P^j\otimes P^{\bj}$ in $\CV_j\otimes
\overline{\CV}_{\bj}$.
If however  $Z_{j\bj}>1 $, $X$ is a linear combination of operators
intertwining the different copies of $\CV_j\otimes  \overline{\CV}_{\bj}$
\eqn\Xform{
P^{(j,\bj;\za,\za')}\ : \ (\CV_j\otimes  \overline{\CV}_{\bj})^{(\za')} 
\to (\CV_j\otimes \overline{\CV}_{\bj})^{(\za)} \quad \za,\za'=
1,\cdots, Z_{j\bj}\ ,}
and acting as $P^j\otimes P^{\bj}$ in each. 
The notation encompasses the case $Z_{j\bj}=1 $.
If $|j,\bn\rangle \otimes |\bj,\bbn\rangle$  denotes an orthonormal 
basis of  $\CV_j\otimes 
\overline{\CV}_{\bj}$ labelled by multi-indices $\bn,\bbn$, 
we may write
\eqn\projP
{P^{(j,\bj;\za,\za')}=\sum_{\bn,\bbn} 
\big(|j,\bn\rangle \otimes |\bj,\bbn\rangle\big)^{(\za)}
\big(\langle j,\bn| \otimes 
\langle\bj,\bbn|\big)^{(\za')}\quad \za,\za'=1,\cdots
Z_{j\bj} \ .}
There are thus $\sum_{j,\bj} |Z_{j\bj}|^2$ linearly independent solutions
of equations \commut.
If these equations are extended to the generators of the full chiral algebra
${\goth A}$, there may be more general  solutions $P_U^{(j,\bj;\za,\za')}$
with, say, $|\bj,\bn\rangle$ replaced by $U\,|\bj,\bn\rangle$,
where $U$ is a unitary operator implementing some automorphism of ${\goth A}$,
a freedom reminiscent to the ``gluing automorphism'' in the boundary
CFT, which  has the effect of changing $(j,\bj)$ to some 
 $(j, \omega(\bj))$.

The $P$'s satisfy 
\eqn\proj
{P^{(j_1,\bj_1;\za_1,\za_1')}P^{(j_2,\bj_2;\za_2,\za'_2)}=
\zd_{j_1j_2}\zd_{\bj_1\bj_2}\zd_{\za_1'\za_2}\, P^{(j_1,\bj_1;\za_1,\za_2')}
\ .}
Note also that they  play here the r\^ ole of  
the Ishibashi states in the problem of boundary conditions in the 
half plane. 
We then 
write the most general linear combination of these basic operators 
as
\eqn\lincom
{X_x= \sum_{j\bj,\za,\za'} {\Psi_x^{(j,\bj;\za,\za')}\over\sqrt{S_{1j}
S_{1\bj}}}\,  P^{(j,\bj;\za,\za')}\ ,}
with $x$ a label taking $n=\sum_{j,\bj}(Z_{j\bj})^2$ values and 
$\Psi$ an a priori arbitrary complex $n \times n$ matrix.  
The denominator $\sqrt{S_{1j} S_{1\bj}}$ is introduced for later
convenience. We shall denote by ${\tCV}$ the set of labels $x$ and  
use the label $x=1$ for the identity operator
\eqn\idenX{X_1:={\rm Id}=\sum_{j\bj,\za} P^{(j,\bj;\za,\za)}\ ,}
for which 
\eqn\Psiid{\Psi_1^{(j,\bj;\za,\za')}= \sqrt{S_{1j} S_{1\bj}}\,\delta_{\alpha
\alpha'}=: \Psi_1^{(j,\bj)}\, \delta_{\alpha \alpha'}  \ .}
Using \proj\ and  the hermitian conjugation properties of the projectors
\vglue-2mm
\eqn\hermit{
( P^{(j,\bj;\za,\za')})^\dagger= P^{(j,\bj;\za',\za)}\, }
\vglue-2mm\noindent
we may compose two such $X$  as
\vglue-1mm\noindent
\eqn\Xcomp{
X_x^\dagger\, X_y= \sum_{j,\bj,\za,\za',\za''}{\Psi_x^{(j,\bj;\za,\za')\, *}\ 
\Psi_y^{(j,\bj;\za'',\za')} \over{S_{1j} S_{1\bj}}}\,
 P^{(j,\bj;\za,\za'')}\ .} 
%
For our purposes, insertion of one or two such $X$ will be sufficient.

\newsec{The consistency equation }

\noindent
As usual, it is convenient to map the cylinder into the complex
plane with coordinate $\zeta$ by  $\zeta=\exp -2i\pi {w\over T}$.
The toroidal domain is mapped into an annulus with identified boundaries
along the circles  $|\zeta|=|\tq|^{\pm 1/2}$
(here $\ttau= 2iL/T$ and $\tq=\exp 2i\pi \ttau$).
One then reexpresses the partition function in terms of Virasoro generators
acting in that plane.
This is a well known calculation \Ca, which is hardly affected by
the insertion of operators $X$ and we find
\eqn\pnfn{
Z_X=  \tr_{\CH_P}( X\, \tq^{L_0-c/24}\, \btq^{\bL_0-c/24})\ , }  
and an analogous formula for the insertion of two $X$.
With the help of 
\eqn\char
{\tr_{\CH_P}( P^{(j,\bj;\za,\za')}
\tq^{L_0-c/24}\, \btq^{\bL_0-c/24})
 = \chi_j(\tq)\, \chi_{\bj}(\btq)\  \delta_{\za\za'}\ ,} 
and of \hermit\
we write the corresponding twisted partition function as
\eqn\Zxy
{Z_{x|y}:=Z_{X_x^\dagger\,X_y}= \sum_{j,\bj\in \CI\atop \za,\za'=1,\cdots,Z_{j\bj}} 
{\Psi_x^{(j,\bj;\za,\za')\, *}
\Psi_y^{(j,\bj;\za,\za')}\over S_{1j}S_{1\bj}} \
 \chi_j(\tq)\, \chi_{\bj}(\btq)\,.
}
In particular, for $x=y=1$, we find
\eqn\modinv{Z_{1|1}= \sum_{j,\bj,\za} 
\chi_j(\tq)\, \chi_{\bj}(\btq)=
\sum_{j,\bj\in \CI}\, Z_{j\bj}\, \chi_j(\tq)\, \chi_{\bj}(\btq)\ ,}
which is, as it should, the modular invariant partition function
describing the system with no twist.
The above discussion may be generalised to the situation where
the underlying chiral algebra of the CFT is a current algebra
of generators $J$ and level $k$
and where the energy-momentum of the system includes a term coupled
to the Cartan generators: $ T'(w)=T(w) -{2i\pi\over T}
\sum_p \nu_p J^p(w) -{k\over 2} \sum_p \big({2\pi\nu_p\over T}\big)^2$, and
a similar expression for $\bar T$. This modification has been
shown in \BPPZ\ to lead to partition functions involving
{\it unspecialised} characters $\chi(\tq,\nu\ttau)$. 
 Repeating this calculation in the present situation (and real $\nu$)
and choosing properly modified projectors
$P_U$,  changing $(j,\bj)$ to $(j,\bj^*)$  
one recovers the analogues of \Zxy\ and \modinv\ with
the second character replaced by $ \chi_{\bj}(\tq, \nu \ttau)^*$.

Because of the identification of its two ends, the  
cylinder considered above may be mapped into another plane,
with coordinate $z=\exp( \pi  {w\over L})$. The image of the 
fundamental domain in $w$ is an annulus in that plane with 
boundaries along the circles  $|z|=1$ and
$|z|=|q|^{-1}$ identified, with now $q=\exp 2i\pi\tau$, $\tau=-1/\ttau=iT/2L$.
Moreover the fact that 
\commut\ is satisfied implies that the energy momentum $T(w), \bT(\bw)$ 
is well defined on the cylinder and consistent with this identification, 
and that  $T(z),\bT(\bz)$ is thus globally defined in the whole plane. 
On the cylinder, one may also use 
the Hamiltonian corresponding to the $\Re e\, w$-translation operator.
Then the partition function $Z_{x|y}$ is obtained  
as the trace of the corresponding evolution operator in a  Hilbert space
\eqn\hilbertxy
{ \CH_{x|y}=\oplus_{i,\bi\in\CI}\, \tV_{i\bi^*;\, x}{}^y\, \CV_i\otimes
\overline{\CV}_{\bi}\ ,}
where the non negative integer multiplicities $ \tV_{i\bi;\, x}{}^y$ depend
on the twists $x$ and $y$. 
In the trivial case $x=y=1$, they must reduce to 
\eqn\trivial{ \tV_{i\bi^*;\, 1}{}^1=Z_{i\bi}\ .}
We can thus complete the calculation as in the
absence of the $X$ operator(s) and get
\eqnn\Zpxy
$$\eqalignno{
Z_{x|y}&=\tr_{\CH_{x|y}}\, q^{L_0-c/24}\,
 q^{\bL_0-c/24}\,  \cr
&=  
\sum_{i,\bi\in\CI} \tV_{i\bi;\, x}{}^y \ \chi_i(q)\, \chi_{\bi}(q)\ .
&\Zpxy }
$$
For real $q, \nu$ the unspecialised analog of the
 second character can be rewritten as
$\chi_{\bi^*}(q, \nu)^*$ and taking into
account \trivial, the partition function $Z_{1|1}(\tau)$ reduces, up to a
$\nu$-dependent factor, to the modular invariant.

Identifying the two expressions \Zxy\ and \Zpxy\ and using
the modular transformations of (unspecialised) characters in the form
$\chi_j(\tq, \nu \ttau)= e^{i \pi k {\nu^2\over \tau}}\,S_{j^*i}\,
\chi_i(q, \nu )$ we get
\eqn\Cardy
{  \tV_{i\bi;\, x}{}^y=
\sum_{j,\bj,\za,\za'}\, 
{S_{ij}S_{\bi\bj} \over S_{1j}S_{1\bj}}\
\Psi_x^{(j,\bj; \za,\za')}\ \Psi_y^{(j,\bj;\za,\za')\, *}
\,,\qquad i,\bi\in\CI \,.
}
In the last step we have used 
the reality of the l.h.s.
The similarity of this condition with Cardy's equation in the case
of open boundaries \Cab\ is not a coincidence. We shall in fact 
exploit  equation \Cardy\ in a way parallel to  Cardy's \BPPZ. 

To proceed, we make the additional assumption that the 
$\Psi_x^{(j,\bj;\, \za,\za')}$ form a unitary (i.e. orthonormal 
and complete) change of basis  
from the $P^{(j,\bj,\za,\za')}$ to the $X_x$ operators. 
The integer numbers $\tV_{i\bi;x}{}^y$ will be regarded either as the entries
of $|\CI|\times|\CI|$ matrices $\tV_x{}^y$, $x,y\in\tCV$,  or as those
of $|\tCV|\times |\tCV|$ matrices $\tV_{i\bi}$, $i,\bi\in\CI$. 

Following a standard argument, equation \Cardy\ may be regarded as
the spectral decomposition of the matrices $\tV_{i\bi}$ into their
orthogonal eigenvectors $\Psi$ and eigenvalues $S_{ij}S_{\bi\bj}/
S_{1j}S_{1\bj}$. As the latter form a representation 
of the tensor product of two copies of Verlinde fusion algebra, the same
holds true for the $\tV$ matrices:
\eqn\doublefus
{ \tilde{V}_{i_1j_1} \tilde{V}_{i_2j_2}=
\sum_{i_3,j_3} N_{i_1i_2}{}^{i_3} N_{j_1j_2}{}^{j_3}\
\tilde{V}_{i_3j_3}\ .}
Combining \trivial\ with \doublefus, we have in particular
\eqn\Mx{
\sum_{i_3j_3} N_{i_1i_2}{}^{i_3} N_{j_1j_2}{}^{j_3}\, Z_{i_3j_3}=\sum_x
\tV_{i_1j_1^*;\, 1}{}^x \tV_{i_2j_2^*;\,x}{}^1 \ , }
which is the way the matrices $\tV_{ij;\, 1}{}^x=\tV_{i^*j^*;\, x}{}^1$
 appeared originally in
the work of Ocneanu.
As will be explained below, all $\tV_x{}^y$ may be reconstructed
from  the simpler Ocneanu matrices $\tV_1{}^x$.

\newsec{Solutions of \doublefus}
\vbox
{\noindent
For the so-called diagonal theories, for which the bulk spectrum 
is given by $Z_{j\bj}=\delta_{j\bj}$, we know a class of solutions
of \doublefus. In that case, it is natural to identify the set $\tCV$ of
twist labels with the set $\CI$ of representations, since their 
cardinality agrees, and to take 
\eqn\diag{\tV_{ij}=N_i N_j}
understood as a matrix product, in particular 
$\tV_{ij;\, 1}{}^k=N_{ij}{}^k$.
The corresponding $\Psi_x^{(j,j)}$ are just the modular matrix elements
$S_{xj}$.  
 As a second case, consider a non-diagonal theory
with a matrix $Z_{ij} = \delta_{i\zeta(j)}$, where $\zeta$ is 
the conjugation of representations  or some other
automorphism of the fusion rules 
(like the $Z_2$ automorphism in the $D_{2\ell+1}$ cases of $\slh(2)$
theories).  Then $ \tV_{ij} =N_iN_{\zeta(j)}$.}

The simplest non trivial  cases are provided by the $\slh(2)$ theories.
The latter are known to be classified by $ADE$ Dynkin diagrams.
The diagonal $A$ and  the $D_{2\ell+1}$ cases have been just discussed. 
In the other cases with a block-diagonal modular invariant partition 
function ($D_{2\ell},\,E_6,\, E_8$), 
we find that the matrices $\tV_{ij;\, 1}{}^x$ may  
be expressed simply as bilinear combinations 
of the matrices $n_{ia}{}^b$
which give the multiplicities of representations when the RCFT lives 
on a finite width strip or in the upper half-plane \BPPZ: there, the indices 
$a,b,\cdots$ belong to the set $\CV$ of vertices of the Dynkin diagram, 
and are in one-to-one 
correspondence with the possible boundary conditions.
In general we find that the labels $x$ may be taken of the form 
$(a,b,\gamma)$, $a,b\in \CV$, $\gamma$ an extra label,  and 
\eqn\Vtilde{\tV_{ij^*;\, 1}{}^{(a,b, \gamma)}=
{\sum_{c\in T_\gamma}} n_{ic}{}^a n_{jc}{}^b}
with $c$ running over a certain subset $T_\gamma$ of vertices. 
For the $D_{2\ell}$ case, we take $b=1$, $\gamma=0,1$ 
and $T_\gamma$ is the set  of vertices of  $\Bbb{Z}_2$ grading equal 
to $\gamma$.  For the conformal embedding cases $E_6$, resp.  $E_8$,
$b=1,2$, resp $b=1,2,3,8$, the label $\gamma$ is dropped,  
and the range of summation of $c$ is the subset $T\subset {\cal V}$, 
identified with the set of representations of the extended fusion 
algebra, i.e., $T=\{1,5,6\}$, resp. $\{1,7\}$.  (For $D_{2\ell}$, $T=T_0$.) 
Here we are making use of the same labelling of vertices
as in \BPPZ.
Finally, the case $E_7$ requires a separate treatment, as all but one of
the matrices $\tV_1{}^x$  may be represented by a formula similar to 
 \Vtilde, in terms of the $n$ matrices 
of the ``parent theory'' $D_{10}$, see \PZnew\ for more details.  
Eq. \Vtilde\ provides closed 
expressions for the matrices $\tV_{ij;1}{}^x$ already known 
from the work of Ocneanu. It
expresses a relation between twisted (torus) and boundary (cylinder)
partition functions, generalising a well known formula for the 
modular invariant. It is illustrated on Fig. 1. 
See also \Reck\
where the partition functions $Z_{x|y}$ for a diagonal theory, cf. \diag, 
appear computing boundary partition functions for tensor product theories.

Let us illustrate \Vtilde\ with the simplest example  $D_4$. There 
$\tCV$ has 8 elements, but one finds only 5 independent matrices 
\eqnn\Dfour
$$\eqalignno{      \tV_1{}^1=\pmatrix{
 1&0&0&0&1\cr  0&0&0&0&0\cr  0&0&2&0&0\cr  0&0&0&0&0\cr
  1&0&0&0&1\cr}
&\quad \tV_1{}^2
=\pmatrix{
 0&0&0&0&0\cr 1&0&2&0&1\cr 0&0&0&0&0\cr 1&0&2&0&1\cr
   0&0&0&0&0\cr}
\quad \tV_1{}^3= \tV_1{}^4=\pmatrix{
 0&0&1&0&0\cr  0&0&0&0&0\cr 1&0&1&0&1\cr  0&0&0&0&0\cr
   0&0&1&0&0\cr}  \cr       
\tV_1{}^5= \tV_1{}^7= \tV_1{}^8&=\pmatrix{
 0&0&0&0&0\cr  0&1&0&1&0\cr 0&0&0&0&0\cr  0&1&0&1&0\cr
   0&0&0&0&0\cr}
\qquad \tV_1{}^6
=\pmatrix{
 0&1&0&1&0\cr  0&0&0&0&0\cr  0&2&0&2&0\cr  0&0&0&0&0\cr
   0&1&0&1&0 \cr}\ . & \Dfour 
\cr}$$
Here the labels 
 $1,2,3,4$ and $5,6,7,8$ refer respectively to $\gamma =0$ and $\gamma=1$ 
in \Vtilde. 

To any of these $\slh(2)$ cases Ocneanu has associated a graph $\tG$
 with a set of vertices given by $\tCV$. 
These graphs are generated by the pair of adjacency matrices, 
$\tV_{2,1}$ and $\tV_{1,2}$. For example in the $D_{2\ell}$ cases in the basis
used above,  $\tV_{2,1}$
is block-diagonal, with $n_2$ appearing twice in the diagonal, while
in  $\tV_{1,2}$ these two blocks appear off diagonally. Such graphs
have been also  constructed in some higher  
 $n \ge 2$ $\slh(n)$ cases, \refs{\Ocn,\BEK}.

The minimal $c<1$ theories are intimately connected to
the $\slh(2)$ ones, as is well known. For the theory 
of central charge  $c=1-6(g-h)^2/gh$ classified  by the 
pair $(A_{h-1},G)$, with $h$ odd, and $g$ the Coxeter number 
of $G$, the  set $\CI$ of  Virasoro representations $(r,s)$  
 is restricted  by $1\le r\le h-1$, $1\le s\le g-1$ and we choose
$r$ odd. Then the twist labels are of the form $(r,x)$, 
$x$ a twist label of the corresponding $\slh(2)$ theory labelled by
$G$ and 
\eqn\minmod{
\tV_{(rs)(r's');\, 1}{}^{(r'',x)}=N_{rr'}{}^{r''}\
\tV^{(G)}_{ss';\, 1}{}^x\ , \qquad r,r',r''\ {\rm odd}\ ,  }
in terms of the fusion matrices $N_r$ of $\slh(2)_{h-2}$ and of
the $\tV^{(G)}$ matrices of the $G$ case of $\slh(2)_{g-2}$
 discussed above.

\newsec{Examples}

\noindent Some of the twisted partition functions of minimal models have 
been already encountered, and have a simple realisation in the 
corresponding lattice models, in terms of defect  
lines or of twisted boundary
conditions imposed on the lattice degrees of freedom. In 
particular, when the underlying lattice Hamiltonian has
some symmetry under a discrete group, one may use any 
element of this group to twist the boundary conditions along 
a line $\CC$, and the invariance of the  Hamiltonian guarantees
the independence with respect to deformations of $\CC$ \Zub: this 
is the lattice equivalent of the property \commut\ above.  

Consider for example the critical Ising model: this is a diagonal
minimal model described by three representations of the Virasoro 
algebra. Here we depart from our previous conventions and denote 
the representations by their conformal weight, $0$, ${1\over 2}$
and ${1\over 16}$. By the previous discussion, we know that there are 
three possible twists,  whose matrix
 $\tV_1{}^x$ is
given by the ordinary fusion matrix $N_x$.  
The one labelled by $0$ corresponds 
to no twist at all, and the two others lead to a partition function
$Z_{0|{x}}$ which reads  
\eqna\ising
$$\eqalignno{
Z_{0|{1\over 2}}&= |\chi_{0}(\tq)|^2 +|\chi_{{1\over 2}}(\tq)|^2-
|\chi_{{1\over 16}}(\tq)|^2  = \chi_{0}(q)\chi_{{1\over 2}}(q)^*+
{\rm c.c.}+|\chi_{{1\over 16}}(q)|^2 
& \ising a \cr
Z_{0|{1\over 16}}&= |\chi_{0}(\tq)|^2 -|\chi_{{1\over 2}(\tq)}|^2=
\big(\chi_{0}(q)+\chi_{{1\over 2}}(q)\big)\chi_{{1\over 16}}(q)^*+
{\rm c.c.} \ .
& \ising b }$$
\vbox{ \nopagenumbers
Other partition functions are then obtained by fusion in the sense
that 
$Z_{y|z}=\sum_x N_{yx}{}^z Z_{0|x}$. In this case, 
only $Z_{{1\over 16}|{1\over 16}}=Z_{0|0}+Z_{0|{1\over 2}}$
is distinct from the previous ones.  

The physical interpretation of the $\tq$ form of \ising{a} is clear: 
the three primary
operators of the theory are weighted by their ${\Bbb Z}_2$ charge.
This is the well known partition function of the Ising model on which 
periodic boundary conditions are imposed on the spin  in one direction
and antiperiodic ones in the other \refs{\Cab,\Zub}. 
In contrast, 
\ising{b} doesn't seem to have been discussed before.
In general, eq. \minmod\ in the diagonal case $G=A$ 
reproduces for $r''=1$, $x=s''=g-1$, $g$ -- even,
  the $\Bbb Z_2$ twisted 
 partition functions   due to antiperiodic
 boundary conditions in \refs{\Cab,\Zub},  see also \Ruel.  The 
physical meaning and implementation in the lattice model of the
others is less clear and would require some further investigation.  
See, however, reference \OP, where new Boltzmann weights 
that preserve 
Yang-Baxter integrability and commutation of the transfer matrices
 are inserted recovering the diagonal series with 
$\tV_{ij;1}{}^x=N_{ij}{}^x$. 

A similar discussion of the  
3-state Potts model, classified as $(A_4,D_4)$,  follows 
easily from the formulae \Dfour, \minmod\ above.
The resulting ten independent partition functions have been
listed in Table 1: $Z_{1|1}$ is the standard modular invariant, 
$Z_{1|3}$ the one studied
in \refs{\vGR,\Cab,\Zub} 
and denoted ``C'' in \Cab: it corresponds to the assignment 
to each operator of the spectrum of its $\Bbb{Z}_3$ charge, $\omega$
or $\bar\omega$ for the Potts spin and parafermion, $1$ for the
others; $Z_{1|5}$ is what was denoted ``T'' in \Cab.

\vglue -10mm
 \centerline{\epsfxsize=16cm\epsfbox{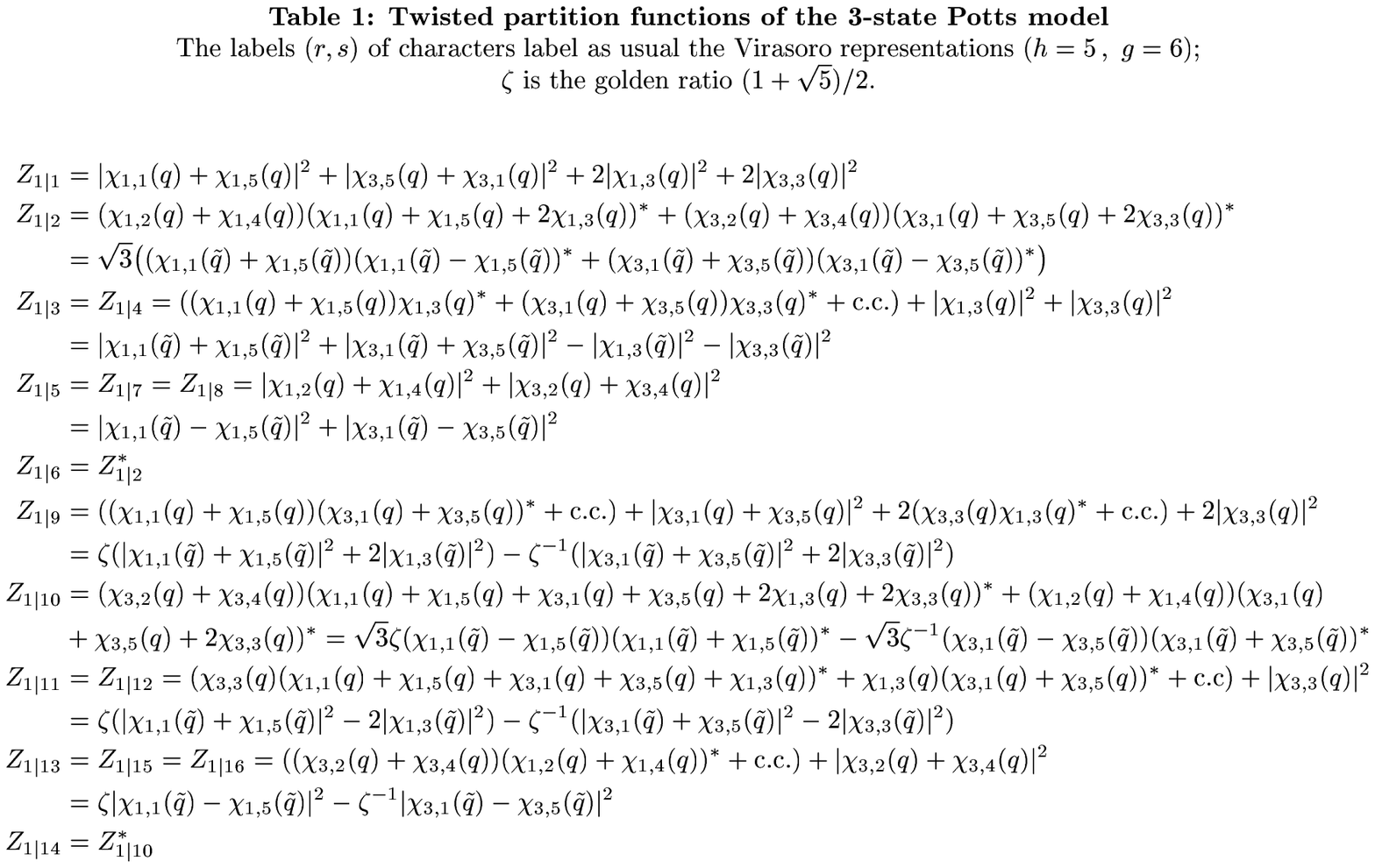}} }

\vbox
{\centerline{\epsfxsize=6cm\epsfbox{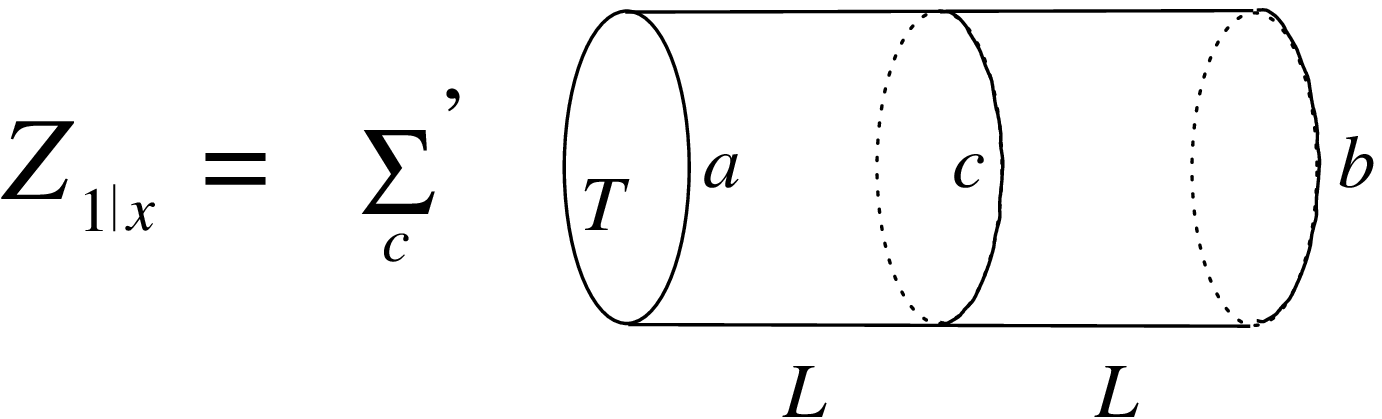}}
\centerline{{\bf Fig 1.} Partition function with the twist $x=(a,b,\gamma)$
as in \Vtilde. }


\vglue -5mm

\newsec{The $\tN$ algebra }

\noindent 
In the diagonal case, formula \diag\  implies that 
 $\tV_y{}^z$ are linear combinations of $\tV_1{}^x$,
 i.e., $\tV_y{}^z=\sum_x N_{yx}{}^z \tV_1{}^x$. This formula
generalises to other cases, the Verlinde matrix being 
replaced by a new nonnegative integer valued 
matrix $\tN_{xy}{}^z$. In terms of the partition functions we have
\eqn\fustw{ Z_{y|z}=\sum_x\, \tN_{yx}{}^z\ Z_{1|x}\ ,}
where
\eqn\Ibl{
\tN_{y x}{}^z=\sum_{j,\bj;\za}\,\sum_{\zb,\zg }\, \Psi_y^{(j,\bj;\,\za,\zb)}\,
{\Psi_x^{(j,\bj;\, \zb,\zg)}\over \Psi_1^{(j,\bj)}}\,
\Psi_z^{(j,\bj;\, \za,\zg)\, *}\,.
}
%
The matrices $\tN_x:=\{\tN_{yx}{}^z\}$ form an associative
algebra $\tN_x\tN_y=\sum_z \tN_{xy}{}^z\tN_z$ 
(``fusion algebra of defect lines''). 
It is noncommutative whenever the corresponding modular invariant
matrix $Z_{j\bj}$ has entries larger than $1$, like e.g. in the $\slh(2)$
$D_{2\ell}$ cases. 
In the commutative cases, \Ibl\ reduces to the spectral representation
of $\tN$.
It is easy to check that in all $\slh(2)$ cases, $\tN_{y x}{}^z$ in
\Ibl\ are indeed non negative integers. This holds true in general
and  finds a natural 
explanation in the framework of the subfactor theory \refs{\Ocn,\BEK}.  
\smallskip
The representations of this fusion algebra are labelled by $(j,\bj)\,, $ 
such that 
$Z_{j\bj}\ne 0\,,$ and appear with multiplicity $Z_{j\bj}$,
i.e. they are in one-to-one correspondence with  
the physical spectrum $(j,\bj;\, \alpha)$ of the bulk theory. 
It turns out that a subset of the structure constants
of the  associated (commutative) algebra  `dual' to the
$\tN$ - algebra  relates to the squared moduli
of the OPE coefficients
of the physical (local) fields, \PZtmr, \PZnew. Thus
all the information about the  bulk theory is encoded in
the eigenvector matrices $\Psi$ of the Ocneanu graphs $\tG$.
We recall that the graphs $G\,, \tG$ and the various multiplicities
-- the sets of integers $N_{ij}{}^k\,, n_{ja}{}^b\,, \tV_{ij;x}{}^y\,, 
\tN_{xy}{}^z\,, \tilde{n}_{ax}{}^b$ --
 are related to the  existence  of a quantum symmetry of the CFT, 
the Ocneanu ``double triangle algebra'', \refs{\PZtmr,\PZnew}, studied 
in the more mathematical language of subfactors in \refs{\Ocn,\BEK}.
The last of the above multiplicities,
 $(\tilde{n})_{ax}{}^b$, furnishes a representation of the $\tN$  algebra,
}  
\eqn\tnm{
\tilde{n}_{ax}{}^b=\sum_{j,\;\za, \zb }\, \psi_a^{(j\,,\za)}\,
{\Psi_x^{(j, j;\, \za,\zb)}\over \Psi_1^{(j,j)}}\,
\psi_b^{(j\,,\zb)\, *}\, ,
}
where $\psi_a^{(j\,,\za)}$ is the  eigenvector matrix  
which diagonalises the cylinder multiplicities $n_i$.

%

\noindent
Note added: Answering a question of Patrick Dorey --
it is easy to repeat the calculation 
on a cylinder in the presence of  twist operators  and with boundary
states $|a\rangle$ and $\langle b|$  at each end.  In the 
simplest case of one such insertion $X_x^\dagger$,  
one gets a partition function linear in the characters with
 multiplicities given by 
$(n_i\tilde{n}_x)_{a^*}^{\ b^*}$, with the notations of \BPPZ. 
%

\bigskip
\noindent{\bf Acknowledgements }
\par\noindent
Our warmest thanks to Adrian Ocneanu for generous
explanations of his recent work, which instigated the present work.  We 
thank Will Orrick, Paul 
Pearce and Andreas Recknagel for explaining the results 
of their work to us prior to publication and Luis Alvarez-Gaum\'e, 
John Cardy  and C\'esar  Gomez for discussions.
V.B.P. acknow\-ledges partial support of the Bulgarian National
Research Foundation (Contract $\Phi$-643).

\bigskip

\footatend\immediate\closeout\rfile\writestoppt
\baselineskip=14pt\centerline{{\bf References}}\bigskip{\frenchspacing%
\parindent=20pt\escapechar=` \input refs.tmp\vfill\eject}\nonfrenchspacing
\end